\documentclass[oneside]{elsart}
\usepackage[pdftex]{graphicx}
\usepackage{amsmath}
\usepackage{amssymb}
\usepackage{amsfonts}
\usepackage{natbib}
\usepackage{hyperref}

\newcommand\de[1]{\,{\mathrm d}#1}

\newcommand{\bmath}[1]{\mbox{\boldmath$#1$}}
\newcommand{\tenss}[1]{\bmath{#1}}                   % A second-order tensor
\newcommand{\tensf}[1]{\bmath{\mathsf{#1}}}          % A fourth-order tensor
\newcommand{\avgs}[1]{\left\langle #1 \right\rangle} % Spatial average

\newcommand{\oavg}[1]{\left\langle\!\left\langle #1 \right\rangle\!\right\rangle}
\newcommand{\emtrx}[1]{\left[\mathsf{#1}\right]}

\journal{Structural Engineering and Mechanics}
\pubyear{2008}
\firstpage{1}

\begin{document}

\begin{frontmatter}

%%%%%%%%%%%%%%%%%%%%%%%%%%%%%%%%%%%%%%%%%%%%%%%%%%%%%%%%%%%%%%%%
\title{Evaluation of homogenized thermal conductivities of imperfect carbon-carbon textile composites using the Mori-Tanaka method}
\author[mech]{Jan Vorel}
\author[mech,cideas]{and Michal \v{S}ejnoha},
\address[mech]{Department of Mechanics, Faculty of Civil
  Engineering, Czech Technical University in Prague, Th\' akurova 7,
  166 29 Prague 6, Czech Republic}
\address[cideas]{Centre for Integrated Design of Advances Structures,
  Th\' akurova 7, 166 29 Prague 6, Czech Republic}

\begin{abstract}
Three-scale homogenization procedure is proposed in this paper to
provide estimates of the effective thermal conductivities of porous
carbon-carbon textile composites. On each scale - the level of fiber
tow (micro-scale), the level of yarns (meso-scale) and the level of
laminate (macro-scale) - a two step homogenization procedure based
on the Mori-Tanaka averaging scheme is adopted. This involves
evaluation of the effective properties first in the absence of
pores. In the next step, an ellipsoidal pore is introduced into a
new, generally orthotropic, matrix to make provision for the
presence of crimp voids and transverse and delamination cracks
resulting from the thermal transformation of a polymeric precursor
into the carbon matrix. Other sources of imperfections also
attributed to the manufacturing processes, including non-uniform
texture of the reinforcements, are taken into consideration through
the histograms of inclination angles measured along the fiber tow
path together with a particular shape of the equivalent ellipsoidal
inclusion proposed already in~\citep{Skocek:2007:MT}. The analysis
shows that a reasonable agreement of the numerical predictions with
experimental measurements can be achieved.
\end{abstract}

\begin{keyword}
Carbon-carbon composites, multi-scale analysis, Mori-Tanaka method,
optimization, porous materials
\end{keyword}

\end{frontmatter}

\section{Introduction}\label{sec:intro}
%%%%%%%%%%%%%%%%%%%%%%%%%%%%%%%%%%%%%%%%%%%%%%%%%%%%%%%%%%%%%%%%%%%%%%%%%%%%%%%%%%%%%%%%%%%%%%%%%%%%%
Since its introduction the Mori-Tanaka (MT)
method~\citep{Mori:1973:MTM} has enjoyed a considerable interest in
a variety of engineering applications. These include classical fiber
matrix composites~\citep{Benvensite:1987:MTM,Sejnoha:MSMSE:96},
natural fiber systems~\citep{Hellmich:TPM:2005}, or even, although
to a lesser extent, typical civil engineering materials such as
asphalts~\citep{Lackner:JMCE:2005} or cement
pastes~\citep{Smilauer:2006:MBM}. While generally applied to modest
geometries, it has also been demonstrated that the MT method may
reliably assist the engineer to meet the ever growing challenge
associated with the analysis of new highly complicated material
systems such as textile reinforced
composites~\citep{Gommers:1998:MTMATC,Huysmans:1998:PIA}.

%%%%%%%%%%%%%%%%%%%%%%%%%%%%%%%%%%%%%%%%%%%%%%%%%%%%%%%%%%%%%%%%%%%%%%%%%%%%%%%%%%%%%%%%%%%%
\begin{figure}
\begin{center}
\begin{tabular}{c@{\hspace{5mm}}c}
\includegraphics[width=6cm]{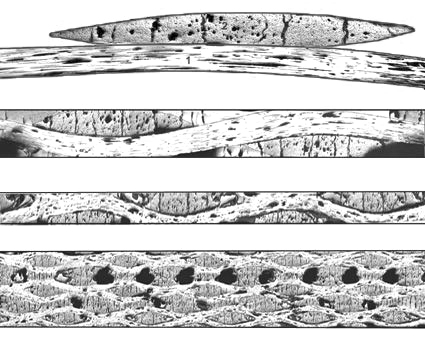}&
\includegraphics[width=8cm]{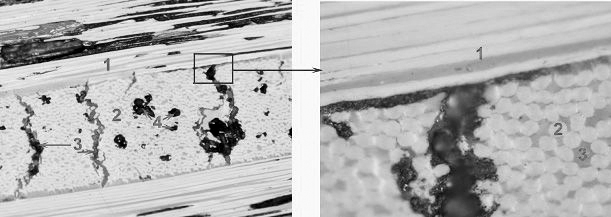}\\
(a)&(b)
\end{tabular}
\caption{Color images of a real composite system: (a) Scheme of multi-scale structural model
(from top - transverse and longitudinal view of fiber tow composite, composite unit cell,
composite lamina, composite plate), (b) Carbon tow microstructure showing major pores
and transverse cracks.}
\label{F1}
\end{center}
\end{figure}
%%%%%%%%%%%%%%%%%%%%%%%%%%%%%%%%%%%%%%%%%%%%%%%%%%%%%%%%%%%%%%%%%%%%%%%%%%%%%%%%%%%%%%%%%%%%

Recent studies addressing the behavior of imperfect carbon-carbon
(C/C) textile composites~\citep{Skocek:2007:MT} further promote the
use of the MT method as it allows for a direct introduction of
various types of imperfections, e.g. in the fiber tow path
represented for example by histograms of distribution of the
fiber-tow orientation angles~\citep{Koskova:2001:DYWP}. Although
carbon-carbon plain weave composites
%reinforced by mutually interlaced systems of unidirectional carbon fiber tows
belong to a progressive material systems with many applications, a
relevant micromechanics model taking into account most of the
geometrical details is still missing. Several appealing routes have
been already offered to satisfactorily reflect commonly observed
imperfections both in the woven path and through the laminate
thickness developed during the manufacturing
process~\citep{Zeman:2003:RC,Skocek:2007:MT}. The reported works
failed, however, to include one of the most important features of
C/C composites illustrated in Fig.~\ref{F1}~-~the porosity, which in
real systems may exceed 30\% of the overall volume.

A general awareness of the need for incorporating the porous phase
in the predictions of overall response of C/C composites has been
manifested in several recent works. While all microstructural
details were properly identified, the actual analysis was limited to
either unidirectional fiber composites represented here by
individual yarns
~\citep{Tsukrov:MAMS:2005,Tsukrov:CST:2007:I,Tsukrov:CST:2007:II} or
finite element simulations of entire laminate performed in
two-dimensional (2D) environment only~\citep{Tomkova:2008:IJMCE}. An
extension of this topic taking into consideration the characteristic
three-dimensional (3D) structure of C/C textile composites is
presented in this paper. The formulation given here is in the spirit
of multi-scale analysis discussed in~\citep{Tomkova:2008:IJMCE}
combined with the application of the MT method to the prediction of
effective elastic properties of C/C composite presented
in~\citep{Skocek:2007:MT}. However, due to the constant importance
of high-temperature systems with a particular role of C/C composites
we consider in this paper the subject of effective thermal
conductivities.

The paper is organized as follows. General description of the
Mori-Tanaka method in the framework of steady state heat conduction
problem is outlined in Section~\ref{sec:MT}. For an extensive list
of references in this subject the reader is referred
to~\citep{Hatta:1986,Benveniste:JAP:90,Jeong:1998}. Two particular
issues are addressed: solution of the problem of a solitary
ellipsoidal inclusion embedded into an orthotropic matrix, see
e.g.~\citep{Chen:1995:AM}, and evaluation of orientation-dependent
average fields. The ordering of the remaining sections follows the
concept of the assumed uncoupled multi-scale homogenization approach
in which the results derived from the homogenization step on a lower
scale are used as an input to the same analysis performed on the
upper scale. Following~\citep{Tomkova:2008:IJMCE} three particular
scales are examined. The level of fiber tow evident from
Fig.~\ref{F1}(b) is treated in Section~\ref{sec:micro} while
Section~\ref{sec:meso} examines various geometrical scenarios
encountered at the level of textile ply, see Fig.~\ref{F1}(a).
Section~\ref{sec:macro} then provides the estimates of the effective
thermal conductivities of the laminate and compares those with the
available experimental measurements. Standard matrix notation is
used throughout the paper.

\section{Mori-Tanaka method}\label{sec:MT}
%%%%%%%%%%%%%%%%%%%%%%%%%%%%%%%%%%%%%%%%%%%%%%%%%%%%%%%%%%%%%%%%%%%%%%%%%%%%%%%%%%%%%%%%%%%%%%%%%%%%%
As illustrated in Fig.~\ref{F1} both micro- (fiber tow) and meso-
(textile ply) scales call for treating at minimum a three phase
composite medium comprising a homogeneous matrix, a certain type of
reinforcement and voids. Combining all three phases in a single
homogenization step appears, however, rather inadequate owing to a
considerable difference in both size and shape between the
reinforcements and voids. Therefore, a two step homogenization
approach is adopted in this paper. In particular, the effective
properties of a composite aggregate are first found in the absence of
pores followed by the second homogenization step in which the porous
phase is introduced into a new homogenized matrix.

\begin{figure}
\begin{center}
\includegraphics[width=5cm]{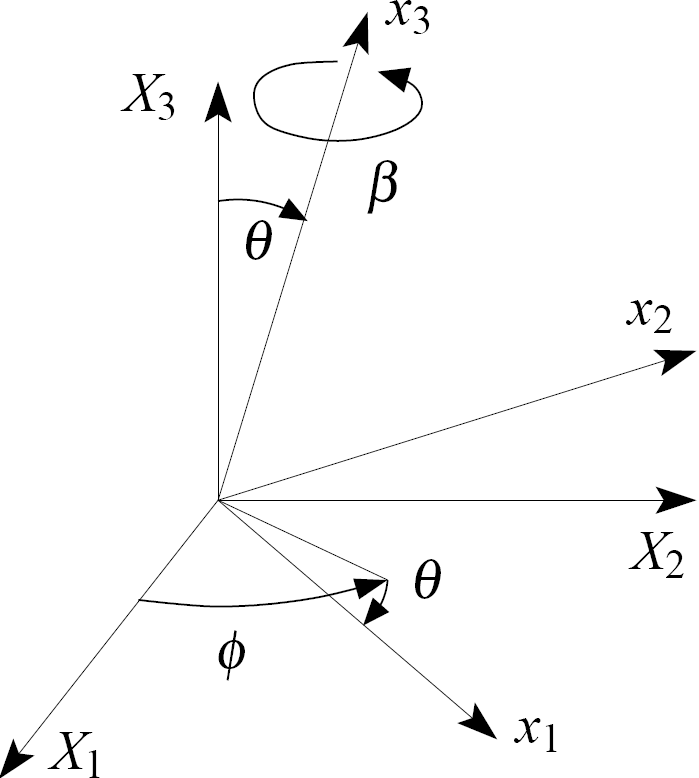}
\end{center}
\caption{Local coordinate system and definition of the Euler angles.}
\label{F2}
\end{figure}

It is therefore sufficient to consider a two-phase composite medium
with the heterogeneities (reinforcements or voids) having in general
a certain orientation distribution. Let this composite be subjected
on its outside boundary $\Gamma$ to a homogeneous temperature
boundary condition defined
as~\citep{Benveniste:JAP:90,Tomkova:2008:IJMCE}
\begin{equation}
\theta(\tenss{X}) = \tenss{H}^{\sf T}\tenss{X}\hspace{1cm}\text{on}\;\;\Gamma,
\end{equation}
where $\tenss{H}$ represents the macroscopically uniform temperature
gradient vector and $\tenss{X}$ are the components of the assigned
(fixed, global) Cartesian coordinate system, Fig.~\ref{F2}. The volume
average of the local constitutive equation written in the local
$\tenss{x}$-coordinate system is then provided by
%
%%%%%%%%%%%%%%%%%%%%%%%%%%%%%%%%%%%%%%%%%%%%%%%%%%%%%%%%%%%%%%%%%%%%%
%%%%%%%%%%%%%%%%%%%%%%%%%%%%%%%%%%%%%%%%%%%%%%%%%%%%%%%%%%%%%%%%%%%%%
%
\begin{equation}
\tenss{q}_r^{x} = -\tensf{\chi}_r^{x}\tenss{h}_r^{x},\label{eq:qr}
\end{equation}
where $\tenss{q}_r$ is the volume average of the local heat flux in
the phase $r$ ($r=1,2$ with $r=1$ reserved for the matrix phase) and
$\tensf{\chi}_r$ is the corresponding thermal conductivity matrix
[Wm$^{-1}$K$^{-1}$]. Also note that $\tenss{x}\equiv\tenss{X}$ for the
matrix phase.

Next, in the context of the Mori-Tanaka method, consider a certain
auxiliary transformation problem where a single heterogeneity in an
infinite matrix is replaced by an equivalent inclusion of the same
shape and orientation but having the material properties of the
matrix phase. In the Mori-Tanaka mean field theory the mutual
interaction of individual heterogeneities is taken into account by
loading such a configuration at infinity by the average temperature
gradient in the matrix $\tenss{h}_1$, see
e.g.~\citep{Benvensite:1987:MTM,Benveniste:JAP:90} for further
details. The local temperature gradient in the inclusion is then
expressed as
\begin{equation}
\tenss{h}_2^{x} = \tenss{h}_1^{x}+(\tensf{S}\tenss{h}^{*})^{x},\label{eq:h2}
\end{equation}
where $\tenss{h}^{*}$ represents a certain transformation
temperature gradient introduced in the homogeneous, generally
anisotropic, matrix to give the same local fluxes as in the
composite. Note that the second order tensor $\tensf{S}$ and the
vector $\tenss{h}^{*}$ are analogous to the Eshelby tensor and
transformation strain known from the solution of the Eshelby
equivalent inclusion problem in elasticity~\citep{Eshelby:1957:SFI}
(hereafter, we shall quote the ``Eshelby problem'' whenever
referring to a general transformation inclusion problem). Recall
that in the MT scheme the $\tensf{S}$ tensor is a function of the
matrix material parameters and the shape of the inclusion. Combining
Eqs.~\eqref{eq:qr} and~\eqref{eq:h2} for the heterogeneity and
equivalent inclusion then gives
\begin{equation}
\tenss{q}_2^{x} = -\tenss{\chi}_2^{x}\tenss{h}_2^{x}\,=\,-\tenss{\chi}_1^{x}\left(\tenss{h}_2^{x}-(\tenss{h}^{*})^{x}\right),
\end{equation}
so that
\begin{equation}
(\tenss{h}^{*})^{x} = (\tenss{\chi}_1^{x})^{-1}\left(\tenss{\chi}_1^{x}-\tenss{\chi}_2^{x}\right)\tenss{h}_2^{x}.\label{eq:h*}
\end{equation}
Next, substitute Eq.~\eqref{eq:h*} into~\eqref{eq:h2} to get
\begin{equation}
\tenss{h}_2^{x}=\tenss{h}_1^{x}+\tensf{S}^{x}(\tensf{\chi}_1^{x})^{-1}\left(\tenss{\chi}_1^{x}-\tenss{\chi}_2^{x}\right)\tenss{h}_2^{x},
\end{equation}
and finally
\begin{equation}
\tenss{h}_2^{x}=\left[\tensf{I}-\tensf{S}^{x}(\tensf{\chi}_1^{x})^{-1}\left(\tenss{\chi}_1^{x}-\tenss{\chi}_2^{x}\right)\right]^{-1}\tenss{h}_1^{x}
\,=\,\tensf{T}_2^{x}\tenss{h}_1^{x},\label{eq:h2-2}
\end{equation}
where $\tensf{I}$ is the identity matrix and the matrix $\tensf{T}_2$
is referred to as the partial concentration factor.

Suppose that the heterogeneity possess a certain orientation
described by the orientation distribution function $g(\phi,
\theta,\beta)$ with $\phi,\theta$ and $\beta$ being the Euler
angles. A particular form of $g$ for plain weaved composites is
given later in Section~\ref{sec:meso}. In general,
following~\citep{Jeong:1998}, the overall average temperature
gradient for a two-phase composite with an orientation-dependent
inclusion given in the global $\tenss{X}$-coordinate system then
attains the form
\begin{equation}
\avgs{\tenss{h}}=c_1\tenss{h}_1+c_2\oavg{\tenss{h}_2},\label{eq:h-va}
\end{equation}
where $c_r$ is the volume fraction of the phase $r$, $\avgs{~}$ stands
for the volumetric averaging and the double brackets $\oavg{~}$ denote
averaging over all possible orientations. The vector $\tenss{h}_2$ in
Eq.~\eqref{eq:h-va} follows from standard transformation of
coordinates so that
\begin{equation}
\tenss{X}=\tensf{Q}\tenss{x}\hspace{0.5cm}\text{and}\hspace{0.5cm}
\tenss{h}_2=\tensf{Q}\tenss{h}_2^{x}\,=\,\tensf{Q}\tensf{T}_2^x\tensf{Q}^{\sf T}\tenss{h}_1\,=\,\tensf{T}_2\tenss{h}_1.\label{eq:h2-g}
\end{equation}
A specific form of the transformation matrix $\tensf{Q}$ consistent
with Fig.~\ref{F2}\footnote{Note that so-called "$x_2$ convention" is
  used; i.e. a conversion into a new coordinates system follows three
  consecutive steps. First, the rotation of angle $\phi$ around the
  original $X_3$ axis is done. Then, the rotation of angle $\theta$
  around the new $x_2$ axis is followed by the rotation of angle
  $\beta$ around the new $x_3$ axis to finish the conversion.} reads
\begin{equation*}
\tensf{Q}=
\emtrx
{\small{
\begin{tabular}{ccc}
$\cos{\phi}\cos{\theta}\cos{\beta}-\sin{\phi}\sin{\beta}$&
$\sin{\phi}\cos{\theta}\cos{\beta}+\cos{\phi}\sin{\beta}$&
$-\sin{\theta}\cos{\beta}$\\
$-\cos{\phi}\cos{\theta}\sin{\beta}-\sin{\phi}\cos{\beta}$&
$-\sin{\phi}\cos{\theta}\sin{\beta}+\cos{\phi}\cos{\beta}$&
$\sin{\theta}\sin{\beta}$\\
$\cos{\phi}\sin{\theta}$&
$\sin{\phi}\sin{\theta}$&
$\cos{\theta}$
\end{tabular}
}}.
\end{equation*}

Next, suppose that the local temperature gradient $\tenss{h}_2$ is
expressed in terms of the prescribed macroscopically uniform
temperature gradient $\tenss{H}=\avgs{\tenss{h}}$ as
\begin{equation}
\tenss{h}_2 = \tensf{A}_2\tenss{H},\label{eq:h2-A2}
\end{equation}
where the matrix $\tensf{A}_2$ is termed the concentration
factor~\citep{Benveniste:JAP:90,Jeong:1998} for the heterogeneity.
Clearly, the orientation average of $\tenss{h}_2$ then follows from
\begin{equation}
\oavg{\tenss{h}_2}=\oavg{\tensf{T}_2}\tenss{h}_1\,=\,\oavg{\tensf{A}_2}\tenss{H},\label{eq:h2-oav}
\end{equation}
which, together with Eqs.~\eqref{eq:h2-2}--\eqref{eq:h2-g} gives
\begin{equation}
\oavg{\tensf{A}_2}=\oavg{\tensf{T}_2}\left[c_1\tensf{I}+c_2\oavg{\tensf{T}_2}\right]^{-1}.
\end{equation}
Combining Eqs.~\eqref{eq:h-va} and~\eqref{eq:h2-oav} further gives
\begin{equation}
c_1\tenss{h}_1=\left[\tensf{I}-c_2\oavg{\tensf{A}_2}\right]\tenss{H}.\label{eq:c1q1}
\end{equation}

In analogy to Eq.~\eqref{eq:h-va}, the volume average of the overall
heat flux is provided by
\begin{equation}
\avgs{\tenss{q}}=c_1\tenss{q}_1+c_2\oavg{\tenss{q}_2}.\label{eq:q-av}
\end{equation}
Next, define an overall conductivity matrix $\tenss{\chi}$ in the
fixed coordinate system $\tenss{X}$ and write Eq.~\eqref{eq:q-av}
as
\begin{equation}
\tensf{\chi}\tenss{H}=c_1\tensf{\chi}_1\tenss{h}_1+c_2\oavg{\tensf{\chi}_2\tenss{h}_2}.\label{eq:chiH}
\end{equation}
Introducing Eqs.~\eqref{eq:h2-A2} and~\eqref{eq:c1q1} into
Eq.~\eqref{eq:chiH} finally provides
\begin{equation}
\tensf{\chi}=\tensf{\chi}_1+c_2\left[\oavg{{\tensf{\chi}}_2\tensf{A}_2}-\tensf{\chi}_1\oavg{\tensf{A}_2}\right].\label{eq:chieff}
\end{equation}

Particular forms of the homogenized matrix $\tensf{\chi}$ will be now
given for individual micromechanics problems.

%%%%%%%%%%%%%%%%%%%%%%%%%%%%%%%%%%%%%%%%%%%%%%%%%%%%%%%%%%%%%%%%%%%%%
\begin{figure}%[ht]
\begin{center}
\begin{tabular}{c@{\hspace{10mm}}c}
\includegraphics[width=5.5cm]{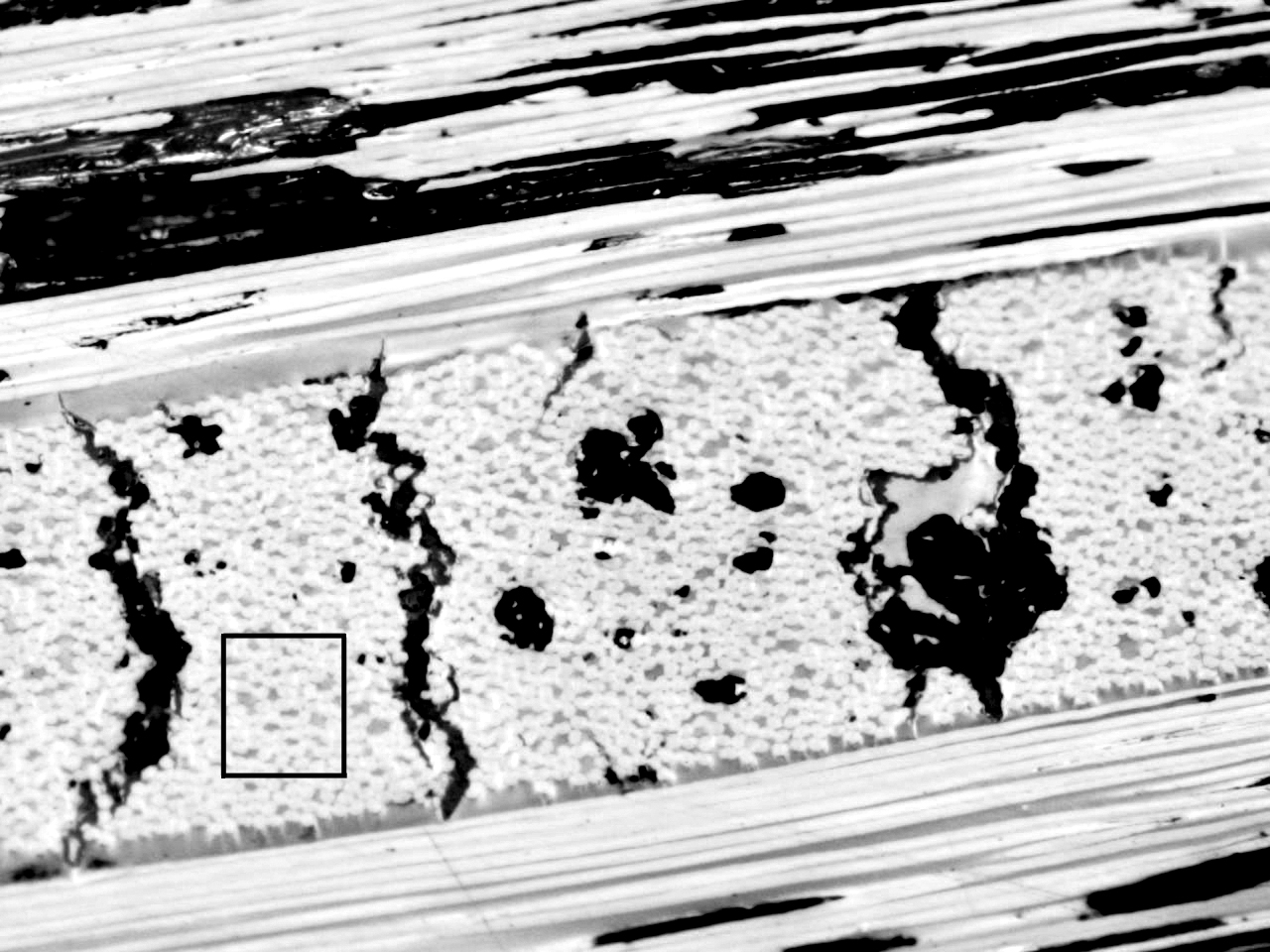}&
\includegraphics[width=5.0cm]{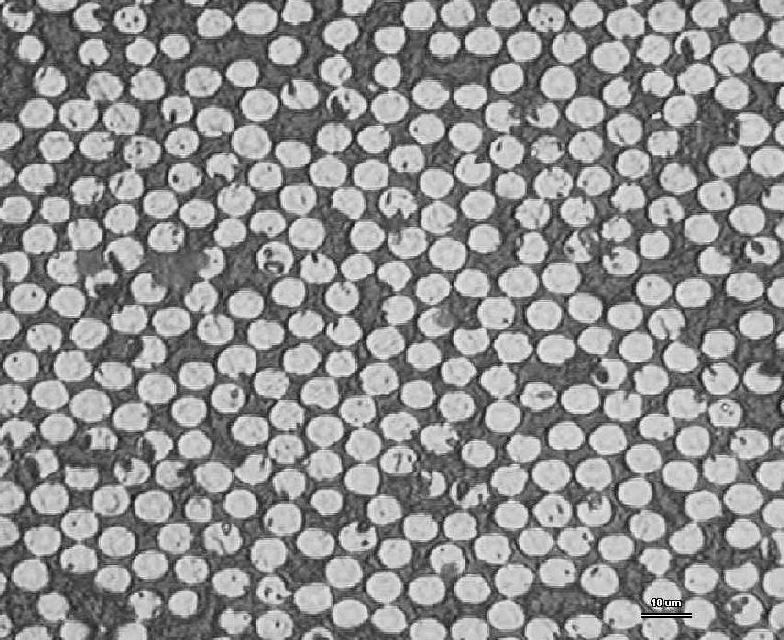}\\
(a)&(b)
\end{tabular}
\caption{Homogenization on micro-scale: (a) fiber tow composite, (b) fiber-matrix composite}
\label{F3}
\end{center}
\end{figure}
%%%%%%%%%%%%%%%%%%%%%%%%%%%%%%%%%%%%%%%%%%%%%%%%%%%%%%%%%%%%%%%%%%%%%

\section{Micro-scale}\label{sec:micro}
%%%%%%%%%%%%%%%%%%%%%%%%%%%%%%%%%%%%%%%%%%%%%%%%%%%%%%%%%%%%%%%%%%%%%%%%%%%%%%%%%%%%%%%%%%%%%%%%%%%%%
In the first step of the proposed multi-scale homogenization scheme
we consider a single filament (fiber tow) of a plain weave carbon
fabric Hexcel 1/1 bonded to a carbon matrix, Fig.~\ref{F3}(a). Each
filament contains about 6000 carbon fibers T800H and significant
amount of transverse cracks and voids resulting in a porosity of
more than 10\%~\citep{Tomkova:2004b,Tomkova:2008:IJMCE}. While the
matrix phase, which essentially corresponds to a glassy carbon, is
assumed isotropic the carbon fiber possess a transverse isotropy
with the value of longitudinal thermal conductivity considerably
exceeding the one in the transverse direction. The phase thermal
conductivities are listed in Table~\ref{T:phases}. Considerable
difference in the size of the two types heterogeneities (fibers and
pores) readily suggests a two step homogenization procedure to
predict the effective properties of the fiber tow as discussed next.

%%%%%%%%%%%%%%%%%%%%%%%%%%%%%%%%%%%%%%%%%%%%%%%%%%%%%%%%%%%%%%%%%%%%%
\begin{table}[ht]
\caption{Material parameters of individual
phases~\citep{Hexcel,Ohlhorst:CCdata}} \label{T:phases}
\bigskip
\centering
\begin{tabular}{|c|c|}
\hline
Material &  Thermal conductivity\\
& [Wm$^{-1}$K$^{-1}$]\\
\hline
Carbon fibers & (0.35, 0.35, 35)\\
Carbon matrix                 & 6.3\\
Voids filled with air         & 0.02\\
\hline
\end{tabular}
\end{table}
%%%%%%%%%%%%%%%%%%%%%%%%%%%%%%%%%%%%%%%%%%%%%%%%%%%%%%%%%%%%%%%%%%%%%

\subsection{Effective conductivities of fiber matrix composites}
%%%%%%%%%%%%%%%%%%%%%%%%%%%%%%%%%%%%%%%%%%%%%%%%%%%%%%%%%%%%%%%%%%%%%%%%%%%%%%%%%%%%%%%%%%%%%%%%%%%%%
Fig.~\ref{F3}(b) shows a representative section of the fiber matrix
composite taken from the fiber tow in Fig.~\ref{F3}(a). Based on our
previous studies~\citep[to cite a
  few]{Zeman:2001:EPG,Sejnoha:2002:OVR,Sejnoha:MCE:2004} such a
composite can be quantified as ergodic, statistically homogeneous with
a random distribution of fibers having the volume fraction of
approximately 50\%. In the MT scheme the effective properties follow
from the solution of an auxiliary problem where an infinite cylinder
of a circular cross-section with semi-axes $\xi_1\rightarrow\infty,\;
\xi_2=\xi_3$ (the $x_1$ axis assumed in the fiber direction) is
embedded into an infinite isotropic matrix.  In this particular case
the effective thermal conductivity matrix given by
Eq.~\eqref{eq:chieff} simplifies as
\begin{equation}
\tensf{\chi}=\tensf{\chi}_1+c_2\left(\tensf{\chi}_2-\tensf{\chi}_1\right)\tensf{A}_2,\label{eq:chieff-micro}
\end{equation}
where
\begin{equation}
\tensf{A}_2={\tensf{T}_2}\left[c_1\tensf{I}+c_2{\tensf{T}_2}\right]^{-1}\hspace{0.5cm}\text{and}\hspace{0.5cm}
\tensf{T}_2\,=\,\left[\tensf{I}-\tensf{S}\tensf{\chi}_1^{-1}\left(\tenss{\chi}_1-\tenss{\chi}_2\right)\right]^{-1}.
\end{equation}
An explicit form of $\tensf{S}$ for this particular case of aligned
circular fibers in an isotropic matrix is available
in~\citep{Hatta:1986}.

\subsection{Effective conductivities of homogenized porous matrix}
%%%%%%%%%%%%%%%%%%%%%%%%%%%%%%%%%%%%%%%%%%%%%%%%%%%%%%%%%%%%%%%%%%%%%%%%%%%%%%%%%%%%%%%%%%%%%%%%%%%%%
Having derived the effective properties of the fiber matrix
composite we proceed with the second homogenization step to account
for the porous phase. As evident in Fig.~\ref{F3}(a), several
distinct shapes of voids can be identified. It is certainly out of
the question to treat each void separately. Therefore, in the
present study, they are all combined into a single equivalent
inclusion resembling an elliptic cylinder. Here, the cylinder is
embedded into a transversely isotropic matrix. However, since the
$S_{11}$ component of $\tensf{S}$ is equal to zero, the solution of
an elliptic cylinder in an isotropic matrix summarized
in~\citep{Hatta:1986} is again applicable. This results in the same
form of estimate of the effective conductivity matrix $\tensf{\chi}$
as given by Eq.~\eqref{eq:chieff-micro}.

\begin{figure}%[ht]
\begin{center}
\begin{tabular}{c@{\hspace{5mm}}c}
\hspace{-0.5cm}
\includegraphics[width=7.5cm]{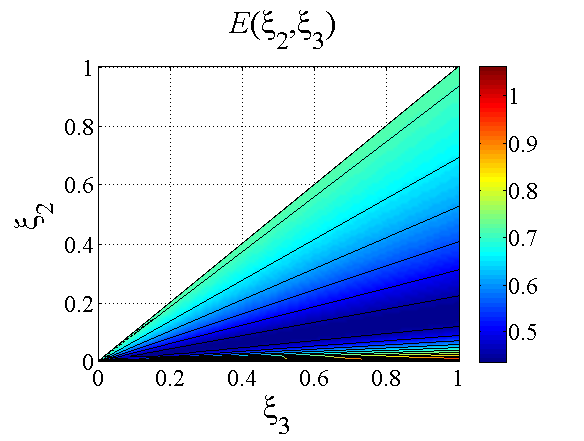}&
\includegraphics[width=7.5cm]{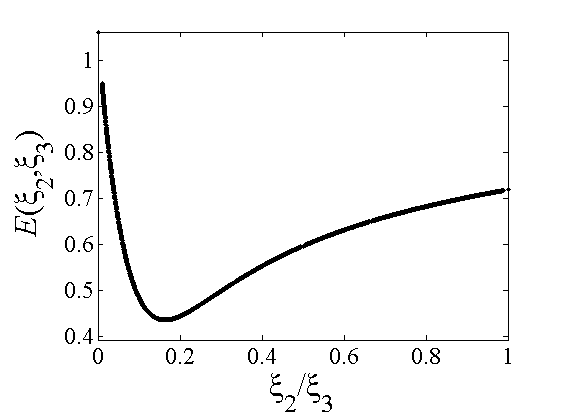}\\
(a)&(b)
\end{tabular}
\caption{Evolution of the objective function as a function of the aspect ratio}
\label{F4}
\end{center}
\end{figure}

Nevertheless, there is still one open problem associated with the
shape of the elliptical cross-section. Clearly, since the equivalent
inclusion represents all possible shapes of voids, it can hardly be
determined directly from the images of real composites such as the
one in Fig.~\ref{F3}(a). Instead, to solve this particular problem,
we exploited the results available from the finite element (FE)
simulations carried out in~\citep{Tomkova:2008:IJMCE}. In
particular, the optimal aspect ratio $\xi_2/\xi_3$
($\xi_1\rightarrow\infty$) of the elliptical cross-section was found
by matching the effective material properties derived from both the
MT method and FE solutions. Fig.~\ref{F4} shows a variation of the
objective function $E$
$$
E(\xi_2,\xi_3)=\left[\sum_{i=1}^3\left(\chi_{ii}^{\rm FEM}-\chi_{ii}^{\rm MT}(\xi_2,\xi_3)\right)^2\right]^{1/2}.
$$
%

%%%%%%%%%%%%%%%%%%%%%%%%%%%%%%%%%%%%%%%%%%%%%%%%%%%%%%%%%%%%%%%%%%%%%
%%%%%%%%%%%%%%%%%%%%%%%%%%%%%%%%%%%%%%%%%%%%%%%%%%%%%%%%%%%%%%%%%%%%%

The resulting effective thermal conductivities provided by the
proposed two-step homogenization scheme are stored in
Table~\ref{T:eff-micro} (Note that the fiber volume fraction
$c_{fiber}$ was estimated from Fig.~\ref{F3}(b), while the volume
fraction of voids $c_{void}=V_{void}/V_{tow}$ stands for the total volume of voids in
the fiber tow in Fig.~\ref{F3}(a)). It also clearly shows a
significance in properly choosing the shape of the cross-section of
the equivalent elliptic cylinder.  Thereby, to make the analysis more
robust, an empirical relation between the observed porosity and
representative equivalent inclusion is needed. This particular topic
enjoys our current research interest.

%%%%%%%%%%%%%%%%%%%%%%%%%%%%%%%%%%%%%%%%%%%%%%%%%%%%%%%%%%%%%%%%%%%%%
\begin{table}[ht]
\caption{Effective thermal conductivities of the fiber tow}
\label{T:eff-micro}
\bigskip
\centering
\begin{tabular}{|c|c|c|c|}
\hline
Material &  \multicolumn{2}{c|}{Equivalent inclusion} & Thermal conductivity\\
\cline{2-3} & fiber ($c_{fiber}$)& void ($c_{void}$)& [Wm$^{-1}$K$^{-1}$]\\
\hline
Fiber-matrix composite & $\infty$,\, 1,\, 1\, (0.55) && 22.09,\, 2.14,\, 2.14\\
\hline
                       & & $\infty$,\, 1,\, 1\, (0.12)& 19.44,\, 1.54,\, 1.54\\
Porous fiber-tow       & & $\infty$,\, 1,\, 10\, (0.12)& 19.44,\, 1.02,\, 1.63\\
(optimal aspect ratio) & $\rightarrow$ & $\infty$,\, 1.6,\, 10\,(0.12)& {\bf 19.44,\, 1.12,\, 1.85}\\
\hline
\end{tabular}
\end{table}
%%%%%%%%%%%%%%%%%%%%%%%%%%%%%%%%%%%%%%%%%%%%%%%%%%%%%%%%%%%%%%%%%%%%%

%%%%%%%%%%%%%%%%%%%%%%%%%%%%%%%%%%%%%%%%%%%%%%%%%%%%%%%%%%%%%%%%%%%%%
\begin{figure}[ht]
\begin{center}
\begin{tabular}{c@{\hspace{5mm}}c}
\includegraphics*[width=70mm]{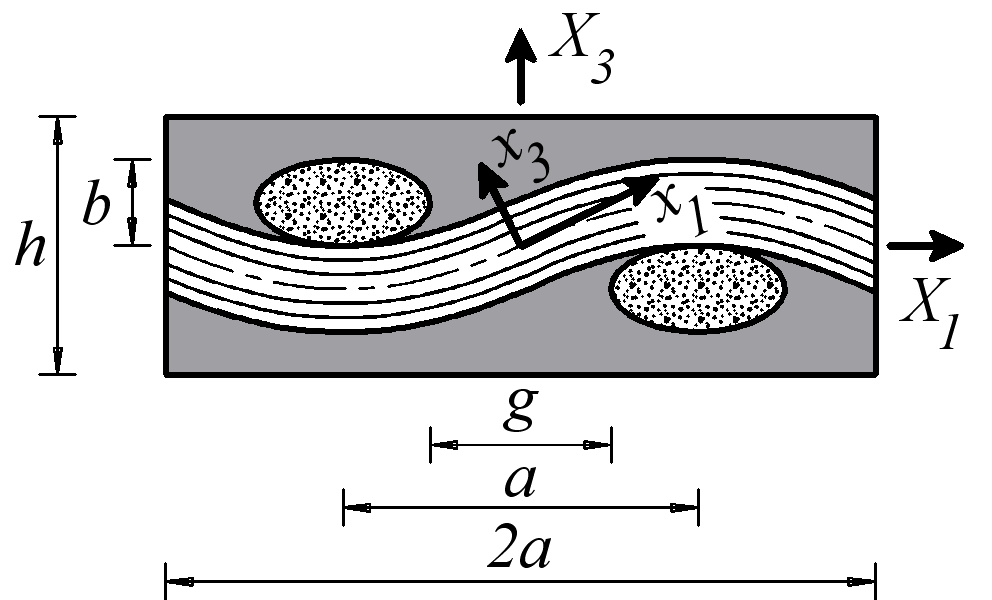}&
\includegraphics*[width=70mm]{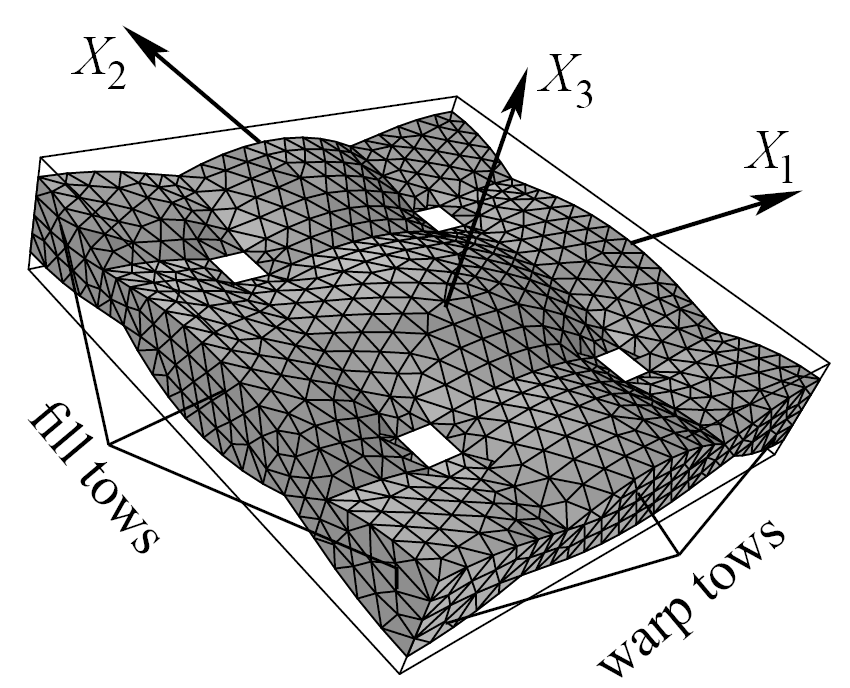}\\
(a)&(b)\\
\includegraphics*[width=65mm]{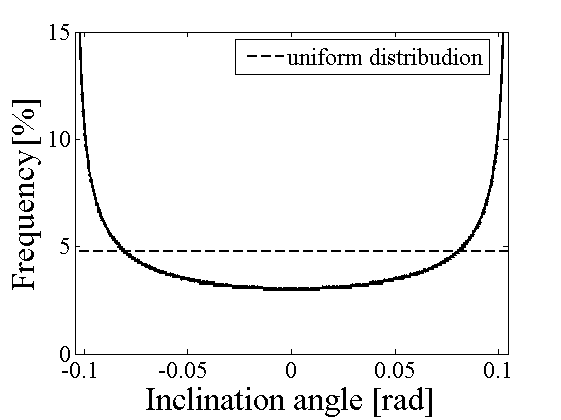}&
\includegraphics*[width=65mm]{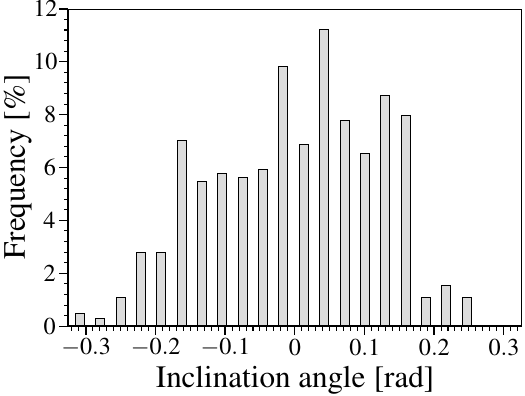}\\
(c)&(d)
\end{tabular}
\caption{Ideal periodic unit cell:
(a) cross-section, (b) three-dimensional view, (c) approximate distribution of inclination angles, (d) example of a real distribution of inclination angles}
\label{F5}
\end{center}
\end{figure}
%%%%%%%%%%%%%%%%%%%%%%%%%%%%%%%%%%%%%%%%%%%%%%%%%%%%%%%%%%%%%%%%%%%%%

\section{Meso-scale}\label{sec:meso}
%%%%%%%%%%%%%%%%%%%%%%%%%%%%%%%%%%%%%%%%%%%%%%%%%%%%%%%%%%%%%%%%%%%%%%%%%%%%%%%%%%%%%%%%%%%%%%%%%%%%%

While unidirectional fiber matrix composites reviewed in the previous
section have been of a general interest since some fifty years ago,
composite systems with a formidable textile texture have received more
attention from both academic and industrial communities only recently.

This section examines, at least from the geometrical point of view,
the most simple representative~-~a plain weave textile composite.
One of the earliest known computational models focusing on actual
geometry of the textile ply is developed in~\citep{Kuhn:1999:MPWCG}.
A three-dimensional view with a typical cross-section are plotted in
Fig.~\ref{F5}(a)(b). The idealized geometry of this model assumes
the centerlines of the warp and fill systems of tows to be described
by a simple trigonometric form
\begin{equation}\label{eq:centerline}
c( x ) = \frac{b}{2} \sin \left( \frac{\pi x}{a} \right).
\end{equation}

Although tempting, a direct application of this model is precluded
by a number of imperfections and irregularities present in real
systems as illustrated in Fig.~\ref{F1}. These include a non-uniform
waviness, mutual shift of individual yarns from layer to layer and
most importantly a non-negligible porosity. Unlike finite element
simulations which enable incorporating most of these imperfections
directly through the formulation of a certain statistically
equivalent periodic unit
cell~\citep{Zeman:2003:RC,Sejnoha:IJES:2007}, the MT method has only
limited means which are, nevertheless, still sufficient when quick
estimates of the effective ``bulk'' response is needed. These will
discussed next in the framework of the two-step homogenization
procedure examined already in the previous section.

\subsection{Effective conductivities of plain weave textile composite ply}
%%%%%%%%%%%%%%%%%%%%%%%%%%%%%%%%%%%%%%%%%%%%%%%%%%%%%%%%%%%%%%%%%%%%%%%%%%%%%%%%%%%%%%%%%%%%%%%%%%%%%
Consider a simple plain weave fabric ply in the absence of porous
ply. At this level, the carbon fiber tow is treated as a homogeneous
phase with known material properties bonded again to an isotropic
carbon matrix. In order to address the influence of various
geometrical flaws, the approach proposed in~\citep{Skocek:2007:MT}
is adopted. This involves:
\begin{enumerate}
\item determination of an ideal geometrical model to asses the volume fraction of the fiber tow
\item determination of the optimal shape of an equivalent ellipsoidal inclusion substituting the fiber tow in the solution of the Eshelby problem
\item proper evaluation of orientation dependent quantities from Eq.~\eqref{eq:chieff} to account for a non-uniform waviness along the fiber tow path
\end{enumerate}

\subsubsection{Ideal geometrical model - Fig.~\ref{F5}(b)}
%%%%%%%%%%%%%%%%%%%%%%%%%%%%%%%%%%%%%%%%%%%%%%%%%%%%%%%%%%%%%
The three-dimensional geometrical model adopted in the present
work~\citep{Kuhn:1999:MPWCG} is characterized by four parameters:
the tow wavelength $2a$, the tow height $b$, tow spacing $g$ and the
layer thickness $h$. To formulate one particular ``ideal''
representative, a tedious image analysis of a number of sections of
a real textile ply such as the one in Fig.~\ref{F1}(a$_1$) was
carried out. The averages of the basic geometrical parameters
presented in Table~\ref{T:PUC_parameters} were used to construct the
required geometrical model.

%%%%%%%%%%%%
\begin{table}[ht]
\caption{Quantification of PUC1 parameters~\citep{Tomkova:2004a}}
\label{T:PUC_parameters} \centering
\begin{tabular}{|c|c|c|c|c|}
\hline
Statistics $\left[\mu m \right]$ & $a$ & $h$ & $b$ & $g$ \\
\hline Average & 2250 & 300 & 150 & 400 \\ \hline Standard deviation
& 155 & 50 & 20 & 105 \\ \hline
\end{tabular}
\end{table}
%%%%%%%%%%%%

\subsubsection{Optimal shape of the equivalent ellipsoidal inclusion}
%%%%%%%%%%%%%%%%%%%%%%%%%%%%%%%%%%%%%%%%%%%%%%%%%%%%%%%%%%%%%%%%%%%%
An extensive numerical study was performed in~\citep{Skocek:2007:MT}
to conceive how the Mori-Tanaka predictions are influenced by a
``random'' deviation of basic geometrical parameters of real systems
from their ideal representative introduced in the previous section.
The results revealed a certain correlation between the model
parameters and ``optimal'' shape of an equivalent ellipsoidal
inclusion characterized by three semi-axes $\xi_1,\,\xi_2,\,\xi_3$.
When setting $\xi_1=1$ (recall that the Eshelby solution depends
only on the mutual ratio of the ellipsoid semi-axes), it was
concluded that the $\xi_2$ parameter is strongly correlated with
$g/a$ ratio, while it is almost independent of $b/a$ value. An
analogous trend could be observed between $\xi_3$ and $b/a$
parameter. This led to following semi-empirical formulas employed
also in this study
\begin{equation}\label{eq:Eshelby_heuristics}
\xi_2 \approx 1 - \frac{3g}{a},\hspace{5mm}
\xi_3 \approx \frac{1}{10} - \frac{b}{3a},
\end{equation}
where the necessary parameters $a, b, g$ are available from
Table~\ref{T:PUC_parameters}.

Remind, however, that these relations were originally derived to match
effective elastic properties. To either confirm or displace their
validity in the solution of heat conduction problem we further assumed
an equivalent inclusion in the form of an infinite elliptic cylinder
with elliptical cross-section estimated directly from actual
cross-section of tows in the representative model giving
\begin{equation}\label{eq:Eshelby_measured}
\xi_1 = \infty,\; \xi_2=12.3,\; \xi_3=1.
\end{equation}

\subsubsection{Orientation averaging}
%%%%%%%%%%%%%%%%%%%%%%%%%%%%%%%%%%%%%%%%%%%%%%%%%%%%%%%%%%%%%%%%%%%%
There are three distinct routes available in this study to reflect
the variation of the inclination angle along the fiber tow path. If
we consider directly the simplified geometrical model in
Fig.~\ref{F5}(b), the joint probability density function
$g(\phi,\theta,\beta)$ results from the harmonic shape of the
centerline as described by Eq.~\eqref{eq:centerline}. Applying the
change of variable formula~\citep[Section 33.9]{Rektorys:1994:SOM},
we obtain after some algebra the expression of the probability
density in the form
$$
g( \phi, \theta, \beta ) = \left\{
\begin{array}{cl}
\displaystyle
\frac{2a}{\pi}\frac{1+\tan^2(\theta)}{\sqrt{b^2\pi^2-4a^2\tan^2(\theta)}}
& \mbox{if } \phi = 0, \beta = 0 \mbox{ and }
-\alpha \leq \theta \leq \alpha, \\
0 & \mbox{otherwise,}
\end{array}
\right.
$$
where
$$
\alpha = \arctan\left( \frac{b\pi}{2a} \right).
$$
Assuming simply a uniform distribution of inclination angles the joint
probability density function attains the form
$$
g( \phi, \theta, \beta ) = \left\{
\begin{array}{cl}
\displaystyle
\frac{1}{2\alpha}
& \mbox{if } \phi = 0, \beta = 0 \mbox{ and }
-\alpha \leq \theta \leq \alpha, \\
0 & \mbox{otherwise}.
\end{array}
\right.
$$
Both functions are plotted in Fig.~\ref{F5}(c) for comparison. In this
study, the latter function was adopted for simplicity.  Next, let
$\tensf{D}$ represents an orientation dependent quantity in
Eq.~\eqref{eq:chieff}
\begin{equation}
\oavg{\tensf{D}} = \oavg{{\tensf{\chi}}_2\tensf{A}_2}-\tensf{\chi}_1\oavg{\tensf{A}_2},
\end{equation}
written, for the warp system, as
\begin{equation}\label{eq:warp}
\oavg{\tensf{D}^{\rm warp}}
=
%\frac{2a}{\pi}
\int_{-\alpha}^{\alpha}
g( \phi, \theta, \beta )
\tensf{D}( 0, \theta, 0 )
\de \theta,
\end{equation}
and similarly for the fill system we get
\begin{equation}
\oavg{\tensf{D}^{\rm fill}}
=
%\frac{2a}{\pi}
\int_{-\alpha}^{\alpha}
g( \phi, \theta, \beta )
\tensf{D}( \frac{\pi}{2}, \theta, 0 )
\de \theta.
\end{equation}
Finally, following~\citep{Skocek:2007:MT}, the resulting homogenized
stiffness matrix given by Eq.~\eqref{eq:chieff} then becomes
\begin{equation}\label{eq:woven_mt}
\tensf{\chi}
=
\tensf{\chi}_1
+
\frac{c_2}{2} \left[
\oavg{\tensf{D}^{\rm warp}}
+
\oavg{\tensf{D}^{\rm fill}}
\right]
,
\end{equation}
One may also suggest to model the plain weave fabric as a three-phase
composite with warp and fill systems of yarns being considered as two
distinct phases. The homogenized conductivity matrix then attains a
slightly different form
\begin{equation}\label{eq:woven_mt-3ph}
\tensf{\chi}=\tensf{\chi}_1+\frac{c_2}{2}\left[\oavg{\tensf{\chi}^{\rm
      warp}\tensf{A}^{\rm warp}+\tensf{\chi}^{\rm fill}\tensf{A}^{\rm
      fill}}-\tensf{\chi}_1\oavg{\tensf{A}^{\rm warp}+\tensf{A}^{\rm
      fill}}\right],
\end{equation}
where
\begin{eqnarray}
\tensf{A}^{\rm warp} &=& \tensf{T}^{\rm warp}\left[c_1\tensf{I}+\frac{c_2}{2}\left(\tensf{T}^{\rm warp}+\tensf{T}^{\rm fill}\right)\right]^{-1},
\hspace{8mm}\tensf{T}^{\rm warp}\;=\;\tensf{T}^{\rm warp}(0,\theta,0),\\
\tensf{A}^{\rm fill} &=& \tensf{T}^{\rm fill}\left[c_1\tensf{I}+\frac{c_2}{2}\left(\tensf{T}^{\rm warp}+\tensf{T}^{\rm fill}\right)\right]^{-1},
\hspace{14mm}\tensf{T}^{\rm fill}\;=\;\tensf{T}^{\rm fill}(\frac{\pi}{2},\theta,0).
\end{eqnarray}
However, the differences in predictions provided by
Eqs.~\eqref{eq:woven_mt} and~\eqref{eq:woven_mt-3ph} are, as seen in
Table~\ref{T:eff-meso-1}, negligible.

Improvements when compared to the assumed ideal geometry are
contained in the third route which allows us to introduce the
non-uniform waviness and to some extent also the mutual shift of
individual layers by utilizing histograms of inclination angles
shown in Fig.~\ref{F5}(d).  These are derived from centerlines of
individual fiber tows described in detail
in~\citep{Vopicka:2004:PGV}. Representing the joint probability
density function by these histograms, the contribution to the
effective conductivity matrix, e.g. from the warp direction
Eq.~\eqref{eq:warp}, reads
\begin{equation}\label{eq:warp-hist}
\oavg{\tensf{D}^{\rm warp}}
=
\sum_{i=1}^{m}
p_i \tensf{D}( 0, \theta_i, 0 ),
\end{equation}
where $m$ denotes the number of sampling values. The discrete angles
$\theta_i$ and probabilities $p_i$ follow directly from the image
analysis data. Eleven such histograms associated with several
sections measured along individual plies were considered. The
resulting averages together with the estimates provided by the
simplified distribution functions are summarized in
Table~\ref{T:eff-meso-1}. For the solution of the Eshelby problem of
an ellipsoidal inclusion in an isotropic matrix we refer the
interested reader to~\citep{Hatta:1986,Jeong:1998}.

\subsection{Effective conductivities of homogenized porous matrix}\label{sec:S-orto}
\begin{figure}[ht]
\begin{center}
\begin{tabular}{c@{\hspace{5mm}}c}
\includegraphics[width=6.5cm]{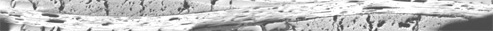}&
\includegraphics[width=6.5cm]{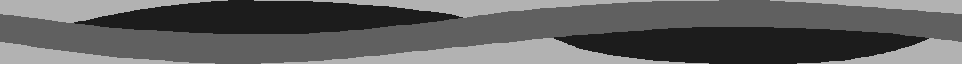}\\
(a)&(d)\\
\includegraphics[width=6.5cm]{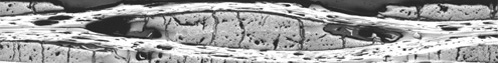}&
\includegraphics[width=6.5cm]{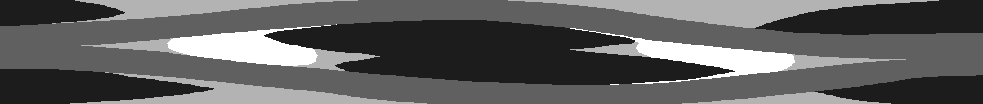}\\
(b)&(e)\\
\includegraphics[width=6.5cm]{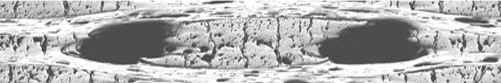}&
\includegraphics[width=6.5cm]{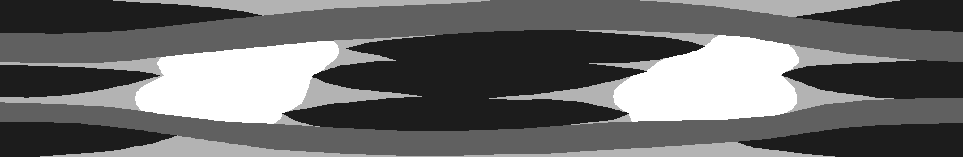}\\
(c)&(f)
\end{tabular}
\caption{Homogenization on meso-scale:
(a)-(b) PUC1 representing carbon tow-carbon matrix composite,
(c)-(d) PUC2 with vacuoles aligned with delamination cracks due to slip of textile plies,
(e)-(f) PUC3 with extensive vacuoles representing the parts with textile reinforcement reduction due to bridging effect in the middle ply}
\label{F7}
\end{center}
\end{figure}

%%%%%%%%%%%%%%%%%%%%%%%%%%%%%%%%%%%%%%%%%%%%%%%%%%%%%%%%%%%%%%%%%%%%%%%%%%%%%%%%%%%%%%%%%%%%%%%%%%%%%
When carefully observing Fig.~\ref{F1}(a$_4$) we identify three more
or less periodically repeating geometries further displayed in
Fig.~\ref{F7}. These segments readily confirm the need for the
proposed two step homogenization procedure as the ideal representative
plotted in Figs.~\ref{F7}(a)(b) (already analyzed in the previous
section) cannot be used to represent the entire composite.  Instead,
the second homogenization step is required to account for the presence
of large vacuoles evident from Figs.~\ref{F7}(c)-(f).

Owing to the orthogonal arrangement of tows in the ideal
(representative) model the new homogenized matrix employed in the
second homogenization step is no longer isotropic. Thereby, the
Eshelby solutions used so far are not directly applicable. Instead,
the $\tensf{S}$ tensor is found by imagining an equivalent ellipsoidal
inclusion in an infinite matrix being orthotropic. The corresponding
Laplace equation governing the steady state heat conduction problem
is provided by
\begin{equation}
\chi_1\frac{\partial^2 \theta}{\partial X_1^2} + \chi_2\frac{\partial^2
\theta}{\partial X_2^2} + \chi_3\frac{\partial^2 \theta}{\partial X_3^2} = 0.
\label{eq:general_Laplace_ortho}
\end{equation}
Introducing the following substitutions
\begin{equation}
X_1 = \sqrt{\chi_1}\tilde{X}_1;\quad X_2 = \sqrt{\chi_2}\tilde{X}_2;\quad
X_3 = \sqrt{\chi_3}\tilde{X}_3,
\end{equation}
allows us to convert Eq.~(\ref{eq:general_Laplace_ortho}) into
\begin{equation}
\frac{\partial^2 \theta}{\partial \tilde{X}_1} + \frac{\partial^2
\theta}{\partial \tilde{X}_2^2} + \frac{\partial^2 \theta}{\partial
\tilde{X}_3^2} = 0, \label{eq:Laplace_ortho}
\end{equation}
which eventually leads to
\begin{equation}
S_{ii} =
\frac{\xi_1\,\xi_2\,\xi_3}{2\sqrt{\chi_1\,\chi_2\,\chi_3}}\int_0^\infty\frac{1}{\left((\xi_i)^2/\chi_i
+s\right) \Delta s}\mathrm{d}s, \label{eq:S_ortho}
\end{equation}
where
\begin{equation}
\Delta s =
\sqrt{\left((\xi_1)^2/\chi_1+s\right)\left((\xi_2)^2/\chi_2+s\right)\left((\xi_3)^2/\chi_3+s\right)}.
\end{equation}
These equations then formally resemble those derived for the case of
an isotropic matrix. The solution of Eq.~\eqref{eq:S_ortho} given in
the form of elliptic integrals is available in~\citep{Jeong:1998}.

\begin{figure}[ht]
\begin{center}
\begin{tabular}{ccc}
\hspace{-0.4cm}
\includegraphics[width=4.63cm]{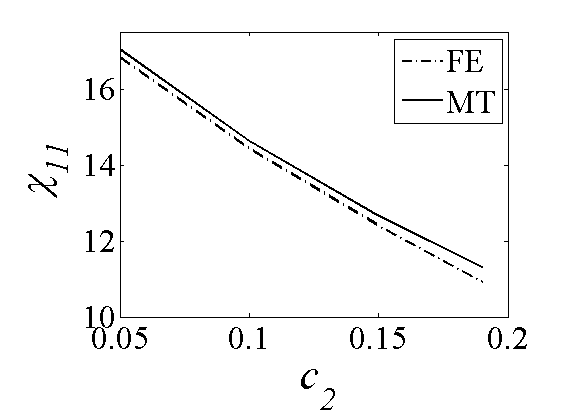} &
\includegraphics[width=4.85cm]{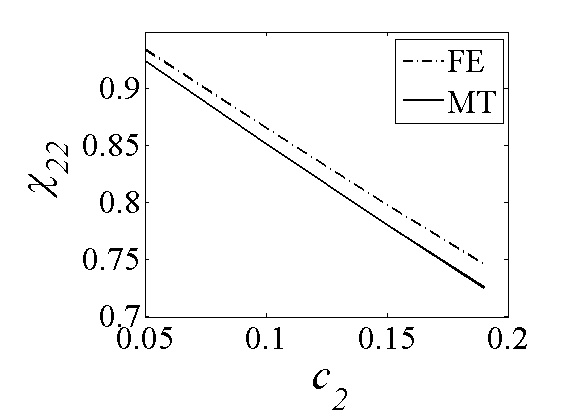} &
\includegraphics[width=4.71cm]{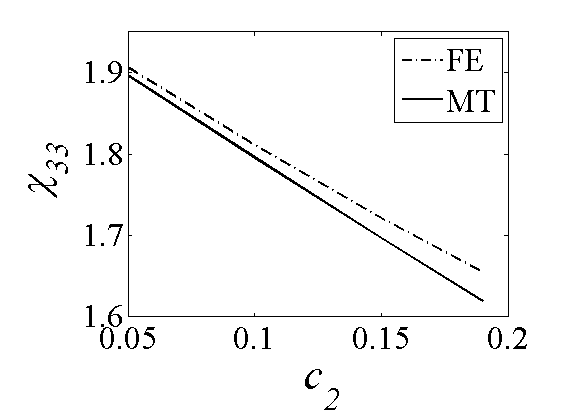} \\
(a) & (b) & (c)
\end{tabular}
\caption{Case study: effective conductivities in [Wm$^{-1}$K$^{-1}$]
  as a function of the volume fraction of the porous phase}
\label{F8:comparison}
\end{center}
\end{figure}

A simple example of an isotropic void (ellipsoidal inclusion with
$\tenss{\xi}=(1,1.5,2)$, $\chi^v=0.2$) surrounded by an orthotropic
matrix ($\chi_{11}^m=20, \chi_{22}^m=1, \chi_{33}^m=2$) was
considered to acknowledge correctness of Eq.~\eqref{eq:S_ortho}.
Fig.~\ref{F8:comparison} compares the MT predictions with the finite
element results found for a hexagonal arrangement of voids under
periodic boundary conditions~\citep{Tomkova:2008:IJMCE}.

%%%%%%%%%%%%%%%%%%%%%%%%%%%%%%%%%%%%%%%%%%%%%%%%%%%%%%%%%%%
%%%%%%%%%%%%%%%%%%%%%%%%%%%%%%%%%%%%%%%%%%%%%%%%%%%%%%%%%%%

With these encouraging results at hand we proceeded with the
analysis of real systems. Unlike the micro-scale, the method of
observation and measuring tools provided by the LUCIA G
software~\citep{LIM} were utilized here to approximate the shape of
individual vacuoles in Figs.~\ref{F7}(c)-(f). Since only
two-dimensional (2D) images were supplied, the voids were assumed to
be well approximated by an oblate spheroid defined in
Table~\ref{T:eff-meso-2} for both types of representative periodic
unit cells in Figs.~\ref{F7}(d)(f).  To estimate the mesoscopic
effective properties the voids were introduced into the homogenized
matrix derived in the 1st homogenization step by combining
Eqs.~\eqref{eq:woven_mt} and~\eqref{eq:warp-hist} (the last two
columns in Table~\ref{T:eff-meso-1}).

A word of caution, however, is appropriate when dealing with 2D images
only. While the shape of the inclusion acquired from 2D images may
play a minor role in final predictions of the effective properties,
the volume fraction of a relevant heterogeneity also estimated from 2D
images may prove much more important. This is documented in
Table~\ref{T:eff-meso-1} listing the predicted effective properties of
a plain weave fabric free of pores for two different volume fractions
of the homogenized fiber tow. The volume fraction 0.53 corresponds to
a representative 3D meso-structure in Fig.~\ref{F5}(b), whereas the
value of 0.78 follows from the corresponding 2D image in Fig.~\ref{F7}(b).

%%%%%%%%%%%%%%%%%%%%%%%%%%%%%%%%%%%%%%%%%%%%%%%%%%%%%%%%%%%%%%%%%%%%%
\begin{table}[ht]
\caption{Effective thermal conductivities of a plain weave fabric without porosity in [Wm$^{-1}$K$^{-1}$]}
\label{T:eff-meso-1}
\bigskip
\centering
\begin{tabular}{|c|c|c|c|c|c|c|c|}
\hline & Fiber tow & \multicolumn{6}{c|}{Thermal conductivity}\\
\cline{3-8}
& volume &
\multicolumn{2}{c|}{Eqs.~\eqref{eq:woven_mt}\&\eqref{eq:warp}}&
\multicolumn{2}{c|}{Eq.~\eqref{eq:woven_mt-3ph}} &
\multicolumn{2}{c|}{Eqs.~\eqref{eq:woven_mt}\&\eqref{eq:warp-hist}}\\
\cline{3-8}
Method & fraction & warp/fill & trans. & warp/fill & trans. & warp/fill & trans.\\
 \hline
MT                                & 0.53 & 8.18 & 3.08 & 7.93 & 3.07 & {\bf 8.14} & {\bf 3.16}\\
\cline{2-8}
Eq.~\eqref{eq:Eshelby_heuristics} & 0.78 & 9.23 & 2.37 & 8.68 & 2.34 & 9.18 & 2.48\\
 \hline
MT                                & 0.53 & 8.35 & 2.94 & 8.26 & 2.92 & 8.31 & 3.02\\
\cline{2-8}
Eq.~\eqref{eq:Eshelby_measured}   & 0.78 & 9.35 & 2.30 & 9.15 & 2.27 & 9.29 & 2.42\\
\hline\hline
3D FEM & 0.53 & {\bf 8.13} & {\bf 3.18} & \multicolumn{4}{c|}{-}\\
\hline
\end{tabular}
\end{table}
%%%%%%%%%%%%%%%%%%%%%%%%%%%%%%%%%%%%%%%%%%%%%%%%%%%%%%%%%%%%%%%%%%%%%

The mesoscopic effective conductivities derived for individual
geometries in Fig.~\ref{F7} are summarized in Table~\ref{T:eff-meso-2}
for both volume fractions of the fiber tow. Note that the results
corresponding to a representative model denoted as PUC1 are
essentially those stored in Table~\ref{T:eff-meso-1} which of course
served as a point of departure for the 2nd homogenization step
performed for models PUC2 and PUC3.

%%%%%%%%%%%%%%%%%%%%%%%%%%%%%%%%%%%%%%%%%%%%%%%%%%%%%%%%%%%%%%%%%%%%%
\begin{table}[ht]
\caption{Effective thermal conductivities of the porous textile plies and laminates in [Wm$^{-1}$K$^{-1}$]}
\label{T:eff-meso-2}
\centering
\begin{tabular}{|c|c|c|c|c|c|c|}
  \hline
  \multicolumn{2}{|c|}{Fiber tow in 1st step} & \multicolumn{3}{c|}{Void} & \multicolumn{2}{c|}{Thermal conductivity}\\
  \hline vol. frac. & shape & PUC & shape & vol. frac. & warp/fill & transverse\\
  \hline &                                                     & PUC2 & $3;3;1$     & 0.07 & {\bf 7.40} & {\bf 2.75} \\
  \cline{3-7} (histograms) & Eq.~\eqref{eq:Eshelby_heuristics} & PUC3 & $1.6;1.6;1$ & 0.15 & {\bf 6.44} & {\bf 2.51} \\
  \cline{2-7} 0.53 &                                           & PUC2 & $3;3;1$     & 0.07 & 7.55 & 2.63 \\
  \cline{3-7} & Eq.~\eqref{eq:Eshelby_measured}                & PUC3 & $1.6;1.6;1$ & 0.15 & 6.57 & 2.40 \\
  \hline &                                                     & PUC2 & $3;3;1$     & 0.07 & 8.32 & 2.18 \\
  \cline{3-7} (histograms) & Eq.~\eqref{eq:Eshelby_heuristics} & PUC3 & $1.6;1.6;1$ & 0.15 & 7.21 & 1.99 \\
  \cline{2-7} 0.78 &                                           & PUC2 & $3;3;1$     & 0.07 & 8.42 & 2.13 \\
  \cline{3-7} & Eq.~\eqref{eq:Eshelby_measured}                & PUC3 & $1.6;1.6;1$ & 0.15 & 7.29 & 1.95 \\
  \hline
\end{tabular}
\end{table}
%%%%%%%%%%%%%%%%%%%%%%%%%%%%%%%%%%%%%%%%%%%%%%%%%%%%%%%%%%%%%%%%%%%%%

%%%%%%%%%%%%%%%%%%%%%%%%%%%%%%%%%%%%%%%%%%%%%%%%%%%%%%%%%%%%%%%%%%%%%
\begin{table}[ht]
\caption{Effective thermal conductivities of the laminate
[Wm$^{-1}$K$^{-1}$]} \label{T:eff-macro}
\bigskip
\centering
\begin{tabular}{|c|c|c|c|c|}
\hline
Method &  \multicolumn{2}{c|}{Fiber tow in 1st step} &\multicolumn{2}{c|}{Thermal conductivity} \\
\cline{2-5}
& shape & volume fraction  & warp/fill & transverse \\
\hline
MT &                                & 0.53 & {\bf 7.26} & {\bf 2.76} \\
\cline{3-5}
(histograms) & Eq.~\eqref{eq:Eshelby_heuristics}  & 0.78 & 8.15 & 2.18 \\
\hline
MT &                                & 0.53 & 7.40 & 2.64 \\
\cline{3-5}
(histograms) & Eq.~\eqref{eq:Eshelby_measured}   & 0.78 & 8.25 & 2.13 \\
\hline\hline
2D FEM~\citep{Tomkova:2008:IJMCE}    & - & 0.78 & {\em 8.67} & {\em 1.66} \\
\hline
Measured~\citep{Kubicar:2002}        & - & - & {\em 10} & {\em 1.6} \\
\hline
\end{tabular}
\end{table}
%%%%%%%%%%%%%%%%%%%%%%%%%%%%%%%%%%%%%%%%%%%%%%%%%%%%%%%%%%%%%%%%%%%%%

\section{Macro-scale}\label{sec:macro}
%%%%%%%%%%%%%%%%%%%%%%%%%%%%%%%%%%%%%%%%%%%%%%%%%%%%%%%%%%%%%%%%%%%%%%%%%%%%%%%%%%%%%%%%%%%%%%%%%%%%%
The final, clearly the most simple, step requires construction of
the homogeneous laminated plate. The stacking sequence of individual
periodic unit cells complies with that observed for the actual
composite sample~\citep[Fig. 2]{Tomkova:2008:IJMCE}, see also
Fig.~\ref{F1}(a$_4$) identifying the PUC1/PUC2/PUC3/PUC1 stacking
sequence. While in-plane conductivities (warp/fill directions) were
found from a simple aritmetic rule of mixture, the out-of-plane
(transverse) conductivity followed from the inverse (geometric) rule
of mixture. The resulting effective thermal conductivities are
available in Table~\ref{T:eff-macro} comparing the MT predictions
and experimental measurements presented in~\citep{Kubicar:2002}. The
FEM results obtained from 2D simulations
in~\citep{Tomkova:2008:IJMCE} are provided for additional
comparison. Note that the highlighted (bolt font) values of thermal
conductivities stored in Tables~\ref{T:eff-micro}--\ref{T:eff-macro}
follow from what we would call an optimal or the most appropriate
approach.

\section{Discussion and conclusions }\label{sec:discussion}
%%%%%%%%%%%%%%%%%%%%%%%%%%%%%%%%%%%%%%%%%%%%%%%%%%%%%%%%%%%%%%%%%%%%%%%%%%%%%%%%%%%%%%%%%%%%%%%%%%%%%
In order to realistically model complex plain weave textile laminates
with three-dimensional, generally non-uniform texture of the
reinforcements and significant amount of porosity we advocate to
consider at least three levels of hierarchy - the level of fiber tow,
the level of yarns and the level of laminate. On each level different
resolutions of microstructural details are considered on individual
scales for the formulation of an adequate representative model. The
desired macroscopic effective properties of the laminate are then
estimated with regard to two basic objectives:
\begin{itemize}
\item to reflect the three-dimensional character of the composite at all scales,
\item to predict the effective conductivities as efficiently as possible.
\end{itemize}
Unlike computationally tedious and extensive 3D finite element
simulations the Mori-Tanaka averaging scheme appears as a reasonable
candidate to comply with both objectives. Not only the fully explicit
format of this method but a simple extension of the Eshelby problem,
at least in the case of the solution of heat conduction problem, to
generally orthotropic reference medium (homogenized composite free of
pores in our particular case) favors this technique.

In this study, the hierarchical character of the analysis is presented
in a totally uncoupled format. Therefore, each level is treated
entirely independently purely upscaling the results from a lower to
higher scale for subsequent calculations. An ``optimal'' procedure,
which attempts to accommodate various sources of imperfections
observed in real systems is accompanied by several modifications
involving mainly the meso-scale.

Based on our previous study of effective elastic
properties~\citep{Zeman:2003:RC,Skocek:2007:MT} it was expected that
at this level the ``best'' estimates of the effective conductivities
would follow from the application of
Eq.~\eqref{eq:Eshelby_heuristics} to determined the shape of an
equivalent inclusion for the fiber tow representation and histograms
of fiber inclination angle to proceed with the orientation averaging
step. Comparing various modifications (different type of inclusion,
ideal path of the fiber tow) suggests, perhaps even intuitively,
almost negligible sensitivity of the solution of the heat conduction
problem to mutual interlacing of individual tows in comparison with
the solution of the elasticity problem. This is mainly attributed to
a relative flatness of the reinforcing yarns in individual plies of
the laminate. In view of this, one may even offer the possibility of
estimating the effective mesoscopic conductivities by simply
assuming two systems of perpendicular fiber tows with no
interlacing, thereby completely avoiding the orientation averaging
step. But bear in mind that such a ``drastic'' simplification can
hardly be generalized and is certainly not acceptable in the case of
elasticity. To discriminate between various approaches is therefore
difficult.

Comparison with experimental measurements is in principle twofold
but also inconclusive. On the one hand it clearly supports the use
of the proposed uncoupled multi-scale approach and the two-step
homogenization scheme on individual scales to arrive at the
predictions of the effective macroscopic thermal conductivities.
Furthermore, at least quantitatively, the Mori-Tanaka method proved
its applicability in the solution of complicated textile composites.
These remarks have already been put forward
in~\citep{Skocek:2007:MT} with regard to the problem of effective
elastic properties. To judge, on the other hand, the pertinence and
reliability of the MT method solely by comparing the predicted and
measured values, which may deny it, is certainly insufficient. While
all deficiencies of the presented homogenization strategy were
openly discussed, errors associated with experimental measurements
were not mentioned and are not available.

In summary, focusing only on the quantitative perspective, the
Mori-Tanaka method combined with popular multi-scale homogenization
approach is viable and presents a suitable and efficient alternative
to periodic homogenization typically based on finite element
analysis.

\section*{Acknowledgment}
%%%%%%%%%%%%%%%%%%%%%%%%%%%%%%%%%%%%%%%%%%%%%%%%%%%%%%%%%%%%%%%%%%%%%%%%%%%%%%%%%%%%%%%%%%%%%%%%%%%%%
The authors would like to thank M. Kop\'{a}\v{c}kov\'{a} for assisting
us with the solution of the transformation problem presented in
Section~\ref{sec:S-orto}. The financial support provided by the
GA\v{C}R Grant No. 106/07/1244 and partially by the research
project CEZ~MSM~684077003 is also gratefully acknowledged.

%%%%%%%%%%%%%%%%%%%%%%%%%%%%%%%%%%%%%%%%%%%%%%%%%%%%%%%%%%%%%%%%%%%%%%%%%%%%%%%%%%%%%%%%%%%%%%%%%%%%%
\bibliographystyle{adfathesis}
\bibliography{liter}

\end{document}